\documentclass[a4paper]{jpconf}
\usepackage{graphicx}

\usepackage{amsmath,amsfonts,amsthm} 

\begin{document}
\title{Pion induced Drell-Yan: the transverse momentum
structure of the pion}

\author{A.~Courtoy}

\address{Instituto de F\'isica, Universidad Nacional Aut\'onoma de M\'exico,
Apartado Postal 20-364, 01000 Ciudad de M\'exico, M\'exico}

\ead{aurore@fisica.unam.mx}

\begin{abstract}
We present an analysis of unpolarized Drell-Yan
pair production in pion-nucleus scattering with a particular focus into the pion dynamics.
The study consists in analyzing the effect of the partonic longitudinal and, especially, transverse distributions of
the pion in a Nambu--Jona-Lasinio (NJL) framework, with Pauli-Villars regularization.
In order to consistently take into account the QCD evolution effects, we have estimated the hadronic scale corresponding to the NJL model's degrees of freedom through a minimization
procedure at NLO: The NLO  evolved pion
distributions have been compared to rapidity differential Drell-Yan cross sections data. That hadronic scale so determined represents the only free parameter in our approach.\\
The NJL transverse momentum PDF, evolved  up to next-to-leading logarithmic accuracy, is then tested against the transverse momentum spectrum
of dilepton pairs up to a transverse momentum of 2 GeV. 
  We found a  fair agreement with available
pion-nucleus data. We find sizable evolution effects on the
shape of the distributions and on the generated average transverse
momentum of the dilepton pair.
\end{abstract}

\section{Introduction}

Drell-Yan (DY) processes  have led to fascinating and challeging physics developments since the '60s. The process is described as follows
\begin{equation}
\label{processDY}
h_1(p_1) \; h_2(p_2)\to \gamma^*(q)+X,
\end{equation}
in which a virtual photon is produced with large 
invariant mass-squared $Q^2$ in the 
collisions of two hadrons at a centre-of-mass energy $s = (p_1 + p_2)^2$, with $p_{1,2}$ the four momentum of hadrons $h_{1,2}$, respectively.
 
 First of all, most fits of parton distribution functions (PDFs) rely on DY data. That is particularily true for pion PDFs whose few fits heavily depend on the pion-induced DY data~\cite{Conway:1989fs}  for which $h_1$ in Eq.~(\ref{processDY}) is a pion and $h_2$ a proton. The latter are the ones that we will consider in the present proceedings.
 
Beyond collinear approaches, the so-called unintegrated cross-sections characterize the spectrum of transverse momentum of the virtual photon, $q_T$. In the kinematical regime in which $q_T$ is of order $\Lambda_{QCD}$, that is  small {\it w.r.t.} $Q$, such an effect is accounted into transverse momentum of the partons through TMDs. The departure from collinearity is here a highly non-perturbative effect, originated in the internal dynamics of the parent hadron.
TMDs have been studied for more than a decade now. While there exist model predictions as well as phenomenological analyses, the implementation of the transverse momentum factorization theorems has added in complexity in globally fitting and phenomenologically determining TMDs. In particular, Semi-Inclusive DIS has been the leading process for such studies, mainly due to experimental tendencies and successes. With the forthcoming pion-induced DY at COMPASS-II, the state-of-the-art theoretical framework will be tested and adjusted from unpolarized  to polarized observables.\footnote{The main physics motivation of such an experiment is to crucially test the universality of the Sivers function, which is expected to have an opposite sign in DY {\it w.r.t} Semi-Inclusive DIS.}
In that context, the structure of the pion plays an important r\^ole. In Ref.~\cite{Ceccopieri:2018nop}, we presented one of the first analyses of the $\pi N$ DY process in terms of the modern TMD formulation. Our method focuses on investigating the DY cross section from the perspective of the dynamics of the pion as embodied by the Nambu--Jona-Lasinio (NJL) model~\cite{ns}.
Complementary studies have been performed in a Gaussian approach in Ref.~\cite{Pasquini:2014ppa} and a phenomenological fit of non-perturbative parameters of the TMD formulation is presented in Ref.~\cite{Wang:2017zym}.

On the other hand, in the complementary kinematical regime, $q_T\sim Q$, perturbative QCD corrections are expected to suffice to explain the transverse momentum spectrum in the regime in which it should apply. However, recent analyses of DY processes find that theoretical predictions based on fixed-order perturbation theory fail to describe Drell-Yan data~\cite{Bacchetta:2019tcu}. This subject is beyond the scope of theses proceedings and our analysis, we however point out that it represents one of the biggest challenge for the next future. 

These proceedings are organized as follows. In Section 2, the pion unpolarized TMD  as obtained in the NJL model in Ref~\cite{ns} is described. Also, a new evaluation of the hadronic scale of the model is presented. In Section 3 we jump to the results with no further details about our choice for the proton structure.

\section{Pion distribution in NJL} 

In this Section we synthetize the most important results of the
the  calculation of pion TMDs in a NJL framework,
with Pauli-Villars regularization of  Ref.~\cite{ns}. Model calculations of meson partonic structure within
this approach have a long story of successful predictions, see {\it e.g.} references in~\cite{CourtoyThesis}.

The unpolarized pion TMD is defined as
\begin{equation}
f^{q/\pi}(x_{\pi},{\bf k}_T; Q_0^2)=\frac{1}{2}\int\frac{d\xi^- d^2\xi_T}{(2\pi)^3}\,e^{-i\left(\xi^-k^+-{\boldsymbol \xi}_T {\bf k}_T\right)}
\langle \pi^+|\bar{\psi}(\xi^-, {\boldsymbol \xi}_T)\gamma^+\frac{1+\tau_3}{2}\psi(0,{\bf 0}_T)|\pi^+\rangle\quad.
\label{defTMD}
\end{equation}
The pion TMD obtained in the NJL framework, here in the chiral limit ---which is
an excellent
approximation to the full result, is given by 
\begin{equation}
f^{q/\pi}(x_{\pi},{\bf k}_T; Q_0^2)= q(x_{\pi}; Q_0^2) T({\bf k}_T)~,
\label{chiral}
\end{equation}
with $q(x_{\pi}; Q_0^2)$ the collinear pion PDF with momentum fraction $x_{\pi}$. In Eq.~(\ref{chiral}), it becomes appearant that the $(x, {\bf k}_T)$ dependences of the pion TMD factorize. Notice that it is no longer the case analytically in the full result. Also, we have assumed that the $Q_0^2$ dependence is carried by the collinear PDF exclusively ; it is not necesarrily legitimate in QCD. 
In the chiral limit,  one has (for $\pi^-$, of interest here) the well-known result
\begin{equation}
\label{ic}
q(x_{\pi},Q_0^2)=d_v(x_{\pi},Q_0^2)=\bar{u}(x_{\pi},Q_0^2)=1\,.
\end{equation}
The function $T$ is given by
\begin{equation}
\label{Tfunction}
T({\bf k}_T)= 
\frac{3}{4 \pi^3} {\left( m \over f_\pi  \right )^2}~
\sum_{i=0,2} \frac{c_i}{k_T^2+m_i^2}=T(k_T) \,,
\end{equation}
which 
satisfies the normalization
\begin{equation}
\label{norm}
\int d^2 {\bf k}_T \, T({\bf k}_T)= 1\,.
\end{equation}

Among the required  properties of the unpolarized TMD obtained in models, we stress that, upon integration over the quark intrinsic 
transverse momentum ${\bf k}_T$,
the pion PDF $q(x)$ is properly recovered with correct
normalization as demonstrate Eqs.~(\ref{chiral},\ref{norm}), and the momentum sum rule is exactly satisfied. The recovery of these properties is due to the fact that NJL is
a field theoretical scheme. Also, the correct
support of the PDF, $0\leq x\leq1$,  arises naturally here. 
Since the result Eq.~(\ref{chiral}) directly results from the definition Eq.~(\ref{defTMD}), the  ${\bf k}_T$ dependence is automatically generated by the NJL dynamics. This is an important feature of the results of Ref. \cite{ns},
not found in other approaches, {\it e.g.} \cite{Pasquini:2014ppa,Wang:2017zym}.
 Since the distribution Eq.~(\ref{Tfunction}) depends only upon $k_T^2$,
its Fourier transform can be cast in the form 
\begin{eqnarray}
S_{NP}^{\pi}(b)&=& \frac{3}{2 \pi^2} {\left( m \over f_\pi  \right )^2}~
\sum_{i=0,2} \int \, dk_T \, k_T \, J_0(b k_T)  
\frac{c_i}{k_T^2+m_i^2}  \nonumber\\
                &=& \frac{3}{2 \pi^2} {\left( m \over f_\pi  \right )^2}~
\sum_{i=0,2} c_i K_0 ( m_i \, b)\,,
\label{bsp}  
\end{eqnarray}
where $K_0$ is the modified Bessel function of the second kind. 
\\
%

In the above definition and results, an additional fundamental parameter is made explicit, {\it i.e.} the hadronic scale of the model $Q_0^2$. Distribution functions as evaluated in models must be matched to the particular low RGE scale at which they mimic the true theory the best. 
Here, the collinear parton distribution obtained within NJL are associated to a specific low 
momentum scale $Q_0^2$ and, 
in order to be used to predict measured quantities,
have to be evolved to higher momentum scales according to
perturbative QCD.
 Such a low scale has been determined previously 
by directly comparing the second moment of the pion PDF evaluated in NJL model with the results from the analysis of Ref.~\cite{Sutton:1991ay}. The procedure gives 
a value of $Q_0^2=0.18$ GeV$^2$ at NLO~\cite{CourtoyThesis,Courtoy:2008nf}\footnote{Other schemes give higher values for the hadronic scale, {\it i.e.} up to $\sim 1$ GeV$^2$~\cite{Noguera:2005cc}.}. 
%
\begin{figure*}
\begin{center}
\includegraphics[scale=0.6]{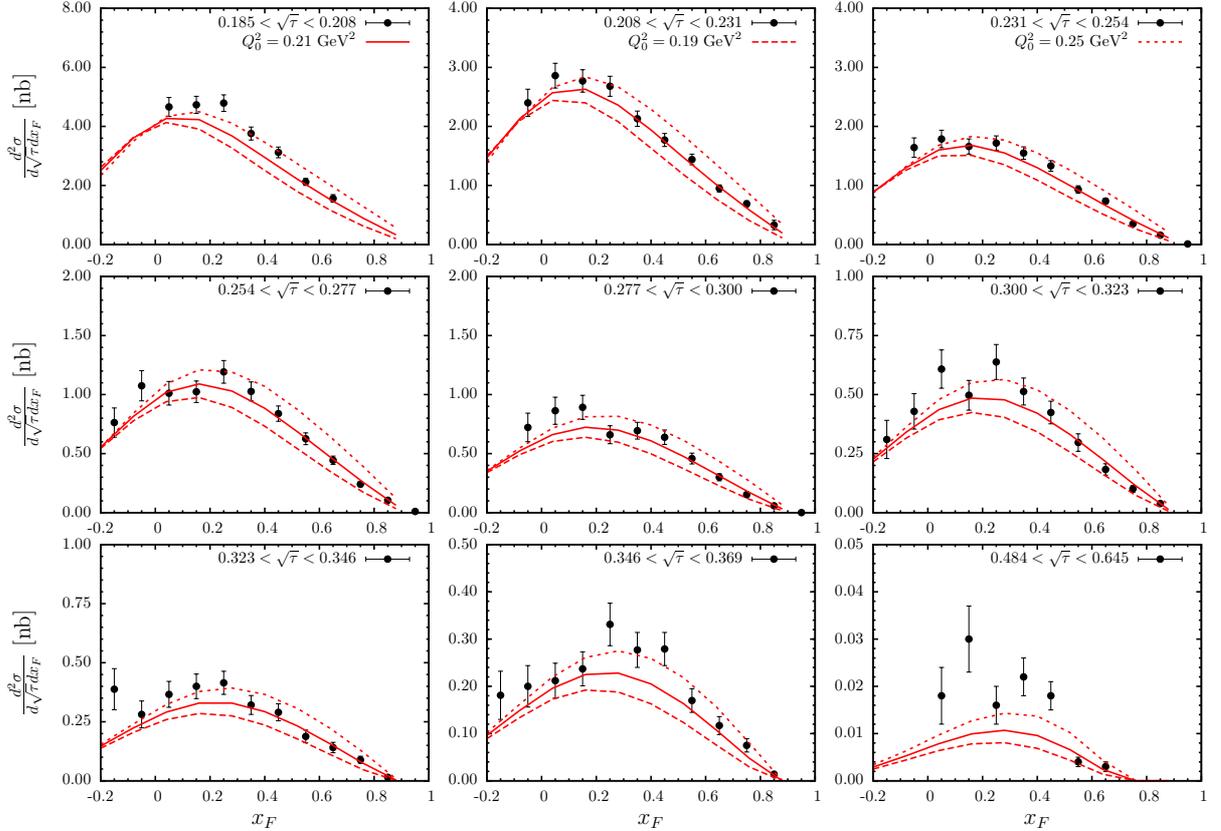}
\caption{Drell-Yan pairs production in $\pi^- W$ collisions. 
Next-to-leading order cross sections obtained by using evolved NJL pion PDFs 
for three values of $Q_0^2$ are compared to data of Ref.~\cite{Conway:1989fs}.}
\label{Fig:2}
\end{center}
\end{figure*}
%

In our analysis, we have used  a different strategy : 
we consider $Q_0^2$ a free parameter of the NJL model which is then fixed with a minimization procedure of the theoretical 
$\pi^- W$ DY cross sections, differential in $\sqrt{\tau}$ and $x_F$,
 against 
the corresponding experimental data~\cite{Conway:1989fs}.
Theoretical cross sections are calculated according to 
\begin{eqnarray}
\label{dsdxf}
&&\frac{d^2 \sigma}{dQ^2 dx_F}=\frac{4\pi \alpha_{em}^2}{9 Q^2 s}\sum_{ij} e_i^2 \int_{x_1}^{1} dt_1 \int_{x_2}^{1} dt_2
\frac{d^2 \hat{\sigma}^{ij}}{dQ^2 dx_F}
f_{i/\pi}(t_1,Q^2) f_{j/p}(t_2,Q^2)\,,
\end{eqnarray}
where the partonic cross sections $d\hat{\sigma}^{ij}$ are calculated at NLO accuracy 
by using the results of Ref.~\cite{Sutton:1991ay}. 
The NJL pion PDFs are evolved to NLO accuracy in the Variable Flavor Number Scheme,
with the initial condition given in Eq.~(\ref{ic}), with the help of the 
QCDNUM~\cite{QCDNUM} evolution code. The QCD parameters 
are those of the NLO CTEQ6M parameterisation~\cite{CTEQ6}. 
In particular we set the NLO running coupling 
to $\alpha_s^{(n_f=5)}(M_Z)=0.118$ at the $Z$-boson mass, $M_Z$. 
Since the data we are comparing to are obtained on a tungsten target, 
we take into account nuclear effects by using nuclear PDFs 
of Ref.~\cite{nCTEQ}.
We have carried out a $\chi^2$ study to establish the hadronic scale of the model that describes  the data the best at NLO in pQCD. Two cases have been considered: an evaluation of the $\chi^2$ for the full range of $x_F$ and another one with a cut $x_F<0.4$, since the NJL model is expected to better reproduce the pion valence distributions, expected to populate the range of large and positive $x_F$ . 
The scales thus determined are
\begin{eqnarray}
Q_{0,\,\mbox{\tiny no cut}}^2=0.212_{-0.012}^{+0.011} \; \mbox{GeV}^2, \quad&& Q_{0,\,\mbox{\tiny cut}}^2=0.209_{-0.009}^{+0.008} \; \mbox{GeV}^2,
\end{eqnarray}
and correspond to a chisquare value of $\chi^2/$d.o.f.$=2.1$ and $1.9$, respectively. The quoted errors correspond to a variation of one unit in $\chi^2$, {\it i.e.} 1-$\sigma$. Those results are compatible with each other.
We will therefore refer to $Q_0^2=0.21$ GeV$^2$,  as the scale associated to the pion NJL model. In Fig.~\ref{Fig:2}, three values for the hadronic scale are depicted for comparison.
It is worth noticing that the results show an acceptable agreement, both in shape and in normalization.

\section{Pion dynamics in DY cross sections}

As mentioned earlier, when $q_T^2$ becomes small compared to $Q^2$, large logarithmic 
corrections of the form of $\alpha_s^n \log^m(Q^2/q_T^2)$ with $0 \leq m \leq 2n-1$
appear in fixed order 
results, being $n$ the order of the perturbative calculation. 
These large logarithmic corrections can be resummed to all
orders by using the Collins-Soper-Sterman (CSS) formalism~\cite{CSS}.
In this limit, the 
cross-section, differential in ${\bf q}_T$, can be written as 
\begin{eqnarray}
\label{CSS}
\frac{d\sigma}{dq_T^2 d\tau dy}&=& \sum_{a,b} \sigma_{q\bar{q}}^{(LO)}  
\int_0^\infty db \frac{b}{2} J_0(b \, q_T)\,S_q(Q,b)\, S_{NP}^{\pi p}(b)  \nonumber\\
&&  \Big[ (f_{a/\pi}\otimes C_{qa})\left(x_1,\frac{b_0^2}{b^2}\right) 
(f_{b/p}\otimes C_{\bar{q}b})\left(x_2,\frac{b_0^2}{b^2}\right)+ q \leftrightarrow  \bar q \Big]\,.
\end{eqnarray}
where $b_0=2 e^{-\gamma_e}$, the symbol $\otimes$ stands for convolution 
and $\sigma_{q\bar{q}}^{(LO)}$ is the leading-order 
total partonic cross section for producing a lepton pair.
The $a,b$ indices run on quark and gluons, and
$J_0(b \, q_T)$ is the Bessel function of first kind. 
The cross section in Eq.~(\ref{CSS}) is also differential in $\tau=Q^2/s$ and $y$, the rapidity of the DY pair. 
Momentum fractions appearing in parton distribution functions 
can be expressed in terms of these variables as
\begin{equation}
x_{1(2)}
= \sqrt{\tau} e^{\pm y},  \,\,\,\,\,\,\,\,  y=\frac{1}{2} \ln \frac{x_1}{x_2}\,.
\label{kine}
\end{equation}
Cross sections differential in $x_F=x_1-x_2= 2 q_\parallel / \sqrt{s}$, the longitudinal momentum of the pair in the hadronic centre of mass system, 
can be obtained from those differential in rapidity $y$, {\it i.e.} $dy     = dx_F / \sqrt{x_F^2 + 4 \tau}$ ; we also
have  that $|x_F|<1-\tau$.
The large logarithmic corrections, exponentiated 
in $b$-space in the Sudakov perturbative form factor, are expressed as
\begin{eqnarray}
\label{sudakov}
S_q(Q,b)
&=&\exp \left\{ -\int_{b_0^2/b^2}^{Q^2} \frac{dq^2}{q^2} 
\left[ A(\alpha_s(q^2)) \;\ln \frac{Q^2}{q^2} + B(\alpha_s(q^2)) \right] \right\}\, .
\end{eqnarray}
The functions $C_{ab}$ and 
$A$, $B$ have perturbative
expansions in $\alpha_s$ and can be found {\it e.g.} in the original paper.

On the other hand, the non-perturbative factor, $S_{NP}^{\pi p}$, contains all the information about the non-perturbative ${\bf k}_T$ behavior. More recent formulations of the CSS formalim directly include the TMD parton densities~\cite{Collins:2014jpa}. Models for hadron's structure yet, as mentioned above, might represent the true theory but at one specific value of the RGE scale, $Q_0$. In that sense, they incorporate a ${\bf k}_T$ ---or equivalently its Fourier conjugate,  as well as the Bjorken $x$, behavior in a fashion that can be resumed as, 
\begin{eqnarray}
f^{q/\pi}(x_{\pi}, b; Q_0^2)&=& q(x_{\pi}; Q_0^2) S_{NP}^{\pi}(b)=q(x_{\pi}; Q_0^2)  \exp\{\ln S_{NP}^{\pi}(b)\}~;
\label{chi_sudakov}
\end{eqnarray}
while the full  TMD parton densities should involve a further $Q^2$ dependence, that is also called   {\it non-perturbative} evolution. Such an evolution has been parameterized in the past, for the proton-proton DY, as
\begin{eqnarray}
\label{fnp}
S_{NP}^{pp}(b)
&=&\exp\{-[a_1 + a_2 \ln (M/(3.2 \,\mbox{GeV})) + a_3 \ln(100 x_1 x_2)] b^2\}\,,
\end{eqnarray}
where $a_1$ plays an equivalent r\^ole to $\ln S_{NP}^{\pi}(b)$ in Eq.~(\ref{chi_sudakov}) and the other parameters, determined  {\it e.g.} in Ref.~\cite{KN05},  reflect a $Q^2$ evolution, {\it i.e.} through the invariant mass $M$ dependence, as well as an unfactorized $(x, {\bf k}_T)$ term. 

Our ansatz here is therefore 
\begin{equation}
\label{prescript_us}
S_{NP}^{\pi p}(b)=S_{NP}^{\pi}(b) \, \sqrt{S_{NP}^{pp}(b)} \,,
\end{equation}
where $S_{NP}^{\pi}(b)$ is given in Eq.~(\ref{bsp}) and the square root on $S_{NP}^{pp}(b)$, given 
in Eq.~(\ref{fnp}).

At very large values of $b$, the perturbative form factor needs to include a {\it taming} through the so-called $b^{\star}$-prescription.
For the process of interest here, it is useful to split the perturbative
form factor in Eq.~(\ref{sudakov}) in a form which allows to use distinct $b_{max}$ on 
the proton and pion side:
\begin{eqnarray}
\label{sudakov_splt}
S_q(Q,b) 
&\equiv& S_q(Q,b_{\star},b_{max}^p,b_{max}^\pi)\nonumber\\
&=&\exp 
\left\{ -\int_{\frac{b_0^2}{b^2_{\star}(b_{max}^p)}}^{Q^2} \frac{dq^2}{2\,q^2} 
\left[ A(\alpha_s(q^2)) \;\ln \frac{Q^2}{q^2} + B(\alpha_s(q^2)) \right] \right\}\,  \nonumber\\
& \times &\exp \left\{ -\int_{\frac{b_0^2}{b^2_{\star}(b_{max}^\pi)}}^{Q^2} \frac{dq^2}{2\,q^2} 
\left[ A(\alpha_s(q^2)) \;\ln \frac{Q^2}{q^2} + B(\alpha_s(q^2)) \right] \right\}~,
\end{eqnarray}
with $b_\star(b,b_{max})=b/\sqrt{1+\Big(\frac{b}{b_{max}}\Big)^2} $ and the respective $b_{max}^p=1.5$GeV$^{-1}$ and the value of $b_{max}^{\pi}$ is adopted such that  $b_{max}^{\pi}=b_0/Q_0=2.44$GeV$^{-1}$. 

\begin{figure*}
\begin{center}
\includegraphics[scale=0.6]{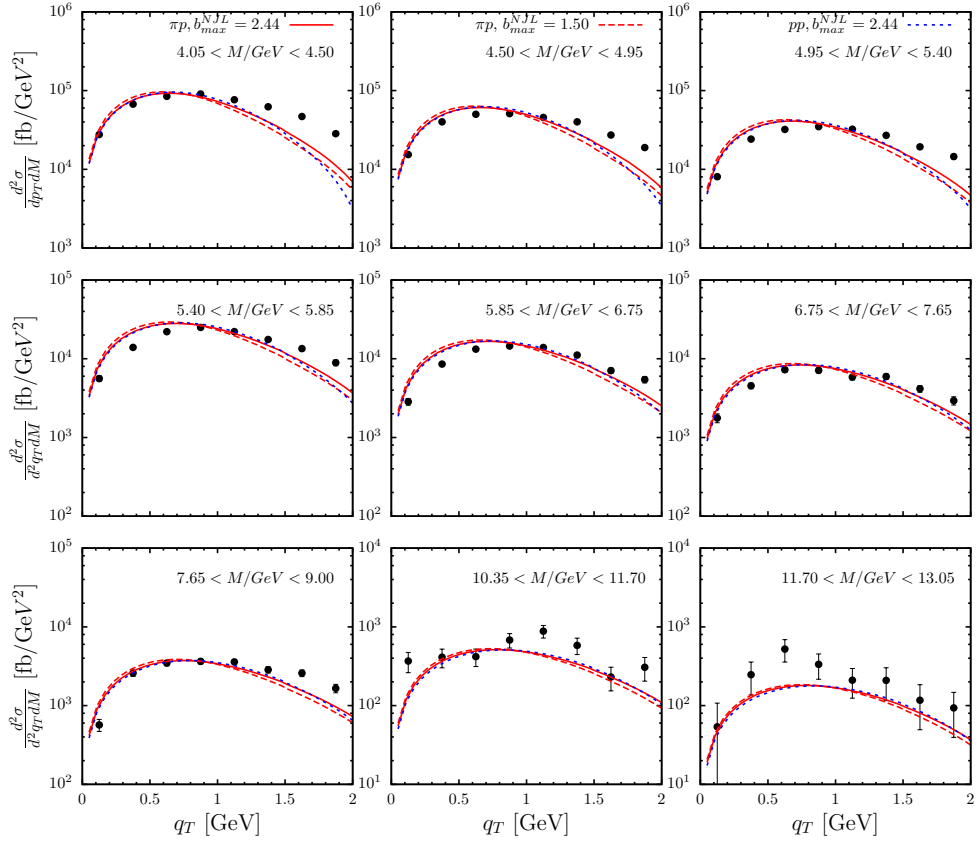}
\caption{Our model results compared to cross sections
in various invariant mass bins of the pair integrated in $0<x_F<1$. 
Data from Ref.~\cite{Conway:1989fs}.}
\label{Fig:4}
\end{center}
\end{figure*}

In Fig.~\ref{Fig:4}, we show the results for the lepton pair $q_T$-spectra of Ref.~\cite{Conway:1989fs}, measured in $\pi W$ collisions. The  cross sections differential in $q_T$ are integrated over $0<x_F<1$ for various $M$ values. The plots range up to $q_T \sim 2$ GeV, range for which the chosen proton description holds~\cite{Ceccopieri:2018nop}. Both red, full and dashed, curves correspond to results using Eq.~(\ref{prescript_us}) with, respectively, the proposed regulator value $b_{max}^{\pi}=2.44$ GeV$^{-1}$ and $b_{max}^{\pi}=1.5$ GeV$^{-1}$ demonstrating the stability of our results at small values of $q_T$. The short-dashed blue curve corresponds to a different ansatz: Eq.~(\ref{fnp}) is used for both hadrons, still with different $b_{max}$ values. At low-$q_T$, the difference between the two ans\"atze is quantitatively small. It supports the hypothesis that the  perturbative evolution,
mainly driven by Eq.~(\ref{sudakov_splt}),  washes away differences  
in the non-perturbative structure, {\it i.e.} gaussian-like with $Q^2$ dependence vs. dynamically generated ${\bf k}_T$ dependence at the hadronic scale $Q_0$. We are tempted to conclude that there is a reduced sensitivity of the data 
to non-perturbative structure. However, this is a first analysis for which no {\it non-perturbative evolution} of the type  Eq.~(\ref{fnp}) has been included, since it is not inherent to the NJL approach used here.
A similar conclusion can be driven for our results for the cross-sections differential  in $q_T$ but integrated over $4<M<8.55$GeV, in the range of $x_F$ where the proton description of~\cite{KN05} is valid.  For higher values of $x_F$, further analyses need to be done before drawing any conclusion.

\begin{figure}[t]
\begin{center}
\includegraphics[scale=0.45]{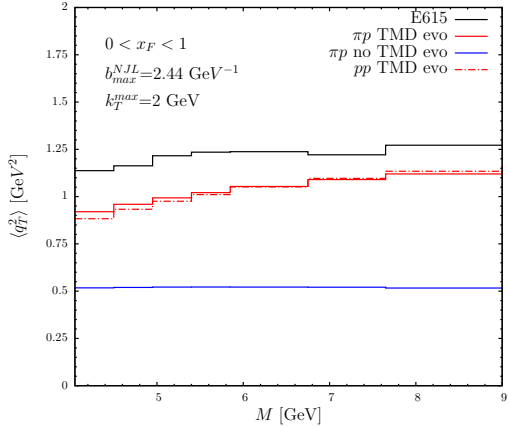}
\caption{Lepton pair average transverse momentum, $\langle q_T^2 \rangle$, as a function of $M$ integrated 
in the range $0<x_F<1$. Averaged values are obtained integrating both predictions and 
the phenomenological parametrization of the data up to $q_T^{max}=2$ GeV.}
\label{Fig:6}
\end{center}
\end{figure}

The average transverse momentum of the pair, $\langle q_T^2 \rangle$, can also be evaluated, as a function of either $x_F$ or $M$. The latter is shown in Fig.~\ref{Fig:6}.  It can be appreciated that the theoretical results undershoot the experiment's, yet reproducing its trend. A slight difference between the Gaussian$+ \, Q^2$-dependence (dot-dashed red curve) and the NJL pion distribution (full red curve)  is seen at  smaller values of the invariant mass of the lepton pair.

\section{Conclusions}

In these proceedings we synthetized the analysis of the DY pair production
in pion-nucleus scattering presented in Ref.~\cite{Ceccopieri:2018nop}.
In that work we tested the outcome of a Nambu--Jona-Lasinio approach for the pion transverse momentum distribution plugged in the CSS framework  at next-to-leading logarithmic accuracy against the differential transverse momentum
spectra of DY pairs produced in 
$pA$ collisions.

The Nambu--Jona-Lasinio approach is a low-energy model for the pion structure, to which is allocated a low hadronic scale, $Q_0$. That hadronic scale is the only free parameter of the model and, in our analysis, it  has been determined  again, with an adapted strategy, confirming its low value, $Q_0^2=0.21$GeV$^2$. 
The agreement found between our pion-nucleus theoretical
cross sections and experimental data is rather successful,
confirming the predictive power
of the NJL model, for both the longitudinal pion parton distributions
and its transverse structure, especially at low values of $q_T$ where no additional contribution to the differential cross-section as described by the CSS formalism, Eq.~(\ref{CSS}), is needed.

Further analyses could include a customized extension of the current approach to include a $Q^2$ dependence. The upcoming data from COMPASS-II on unpolarized target need being understood before turning theorists' attention to the long awaited DY with transversely polarized proton target and its promising access to the Sivers function.

\section*{Acknowledgments}
The author thanks her colleagues and co-authors of the original publication, F.~A.~Ceccopieri, S.~Noguera and S.~Scopetta.  This work is supported in part by DGAPA-PAPIIT IA102418.

\end{document}